\documentclass[usenatbib,usegraphicx]{mn2e}
\usepackage{graphicx}

\title[Globular Clusters and Dwarf Spheroidal Galaxies]{Globular Clusters and Dwarf Spheroidal Galaxies}

\author[Sidney van den Bergh]
       {Sidney van den Bergh$^1$\thanks{E-mail: sidney.vandenbergh@nrc-cnrc.gc.ca}\\
        $^1$Herzberg Institute of Astrophysics, 
        National Research Council, 5071 West Saanich Road, \\
        Victoria, B.C., V9E 2E7, Canada \\}

\begin{document}

\date{Received  }

\pagerange{\pageref{firstpage}--\pageref{lastpage}} \pubyear{ }

\maketitle

\label{firstpage}

\begin{abstract} 
             
Traditionally globular clusters and dwarf spheroidal galaxies have
been distinguished by using one or more of the following criteria: (1) mass, (2) luminosity, (3) size, (4) mass-to-light ratio and (5) spread in metallicity. However, a few recently discovered objects show some overlap between the domains in parameter space that are occupied by galaxies and clusters. In the present note it is shown that ellipticity can, in some cases, be used to help distinguish between globular clusters and dwarf spheroidal galaxies.

\end{abstract}


\section{INTRODUCTION}
 Recent discoveries of exceedingly faint dwarf spheroidal galaxies
have shown quite dramatically that luminosity alone cannot be relied on to distinguish between dwarf spheroidal galaxies and globular clusters.
Interesting examples of faint dwarfs are Bo\"{o}tes II [M$_{v}$ = -3.1] \citep{      wjw07}, Coma [M$_{v}$ = -3.7] \citep{bel07} and Willman 1 [M$_{v}$ = -2.5] \citep{mar07}. All of these galaxies are fainter than the overwhelming majority of globular clusters.

The size of a globular clusters is best described by its half-light
radius R$_{h}$, because this parameter remains almost invariant over $\sim$ 10 cluster relaxation times. The vast majority of globular clusters have R$_{h} <$ 10 pc. However, some well-established globular clusters have quite large radii. Examples are NGC 2419 (R$_{h}$ = 18 pc) and Palomar 14 (R$_{h}$ = 25 pc). On the other hand Willman 1 \citep{mar07}, which has a half-light radius of $\sim$ 20 pc, has generally been regarded as a galaxy.  This overlap in size and luminosity raises deep questions about the nature of the distinction between these two classes of objects. It should be noted that the distinction between galactic nuclei and globular clusters is also somewhat artificial. Over time tidal forces will, for example, strip away much of the stellar population of the Sagittarius dwarf spheroidal, leaving behind only its nucleus the luminous globular cluster M54. \citet{vdbm04} and \citet{macvan05} have suggested that the position of an object in a plot of M$_{v}$ versus log R$_{h}$ might be used to discriminate between globular clusters and the stripped cores of dwarf galaxies, such as $\omega$ Centauri. However, this proposal now seems less attractive than it once did . In particular \citet{vdb07} has recently noted that the brightest red objects in the halo of the elliptical galaxy NGC 5128 appear to form a continuum in the M$_{v}$ versus log R$_{h}$, {\bf indicating that globular clusters and dwarf spheroidals may not be clearly separate and distinct types of objects.  The same point has recently been made about the brightest objects in NGC5128 by \citet{barmby07} and by \citet{rej07}}. Furthermore NGC 2419, which on the basis of its position in the M$_{v}$ versus log R$_{h}$ plane, had been classified as a stripped galaxy core, turns out to have a small metallicity dispersion and lacks a significant population of extra-tidal stars \citep{bellrip07}. Both of these factors militate against the hypothesis that NGC 2419 is actually a stripped galaxy core. On the other hand the stripped core suspect B514 in the Andromeda galaxy does seem to be embedded in a very low surface density dwarf spheroidal \citep{fre07}. In summary it appears that the location of a galaxy in the Mv versus log Rh plot is not always a reliable way of separating globular clusters from the stripped cores of dwarf spheroidal galaxies.

The pioneering investigations of \citet{bab39} and \citet{oor40}
first demonstrated that dark matter provides a significant contribution to the masses of individual galaxies. More recently \citet{mat98} showed that such dark matter becomes more and more dominant as one proceeds to study ever dimmer galaxies. It has therefore become customary to regard the presence of dark matter as the touchstone that allows one to unambiguously distinguish between galaxies and star clusters. Furthermore, it is often difficult and time consuming to obtain velocity dispersions of faint stars in distant dwarf galaxies. The problems outlined above suggest that it might be useful to have additional criteria to help discriminate between dwarf spheroidal galaxies and globular clusters.

\section{ELLIPTICITY AND CLASSIFICATION}

    Table 1 shows the frequency distribution of the ellipticity parameter (a-b)/a , in which a and b are, respectively, the semi-major and semi- minor axes. For Galactic galactic globular clusters the data in the table were taken from the compilation of \citet{har96} that was updated in 2003 at http://physwww.mcmaster.ca/~harris/mwgc.dat. Also given in the table is the distribution of flattening values, drawn from various sources, for those dwarf spheroidal galaxies that have distances $<$ 500 kpc. The individual values of (a-b)/a that were adopted are: CVn II 0.30, Car 0.33, Com 0.50, Dra 0.29, For 0.30, Her 0.67, Leo T 0.0:, Leo I 0.37, Leo II 0.13, Leo IV 0.25, Segue I 0.30, Scl 0.32, Sex 0.35 and UMi 0.66. A comparison between the distribution of globular cluster and dwarf spheroidal galaxy flattenings is listed in Table 1 and plotted in Figure 1. These data show that the the dwarf spheroidal companions to the Galaxy are typically much more flattened than are Galactic globular clusters. A Kolmogorov-Smirnov test shows that the observed difference is significant at $>$99.99\%. A K-S test also shows that the distribution of the flattening values of six M31 dwarf spheroidals \citep{mccirw06} is statistically not distinguishable from that of the Galactic dwarf spheroidals discussed above. The reason for the difference between the the flattening distributions of globular clusters and of dwarf spheroidal galaxies is not yet entirely clear, but might reflect the relative importance of dissipative effects during the evolution of clusters and dwarf galaxies.  {\bf Alternatively the observed flattening of dwarf spheroidal galaxies might reflect the shapes of their original dark matter halos}. Galactic tides might also be important. As \citet{goo97} has pointed out the strength of the tidal field of a parent galaxy may affect the ellipticities of globular clusters. This is so because a strong tidal field might rapidly destroy velocity anisotropies in initially tri-axial rotating globular clusters. However, a possible argument against the importance of tidal effects is that the (a-b)/a values of Galactic globular clusters are not correlated with either their half-light radii or with their concentration indexes. Furthermore tidal fields might not account for all distortions of dwarf spheroidal galaxies. As \citet{coljon07} point out the Hercules dwarf spheroidal, which is located at a distance of 130 kpc, would need to have had a periGalactic distance of $\sim$ 8 kpc to account for its present three-to-one axial ratio.

The large systematic difference in average flattening between dwarf
spheroidal galaxies and globular clusters seems to provide a useful way of distinguishing between galaxies and clusters. All Local Group objects with (a-b)/a $>$ 0.3 appear to be galaxies. Unfortunately projection effects only allow one to draw statistical inferences about the nature of any individual object with (a-b)/a $<$ 0.3. Adopting the criterion that objects with (a-b)/a $ >$ 0.3 are dwarf spheroidal galaxies one finds that four out of 23 of the most luminous ``globular clusters'' surrounding the giant elliptical galaxy NGC 5128 \citep{rej07} might actually be dwarf spheroidal galaxies. It is noted parenthetically that the faint pair of NGC 5128 clusters C141 and C144, have (a-b)/a values of 0.60 and 0.68 respectively \citep{har06}. It would clearly be of considerable interest to investigate this close pair (separation 2.'3) of unusually flattened objects in more detail.
  
Clusters that have been suspected of being the stripped cores of
now defunct dwarf spheroidal galaxies have flattening values that are intermediate between those of typical globular clusters and dwarf spheroidal galaxies. For such objects (a-b)/a = 0.17 in NGC 5139 ($\omega$ Cen), 0.06 for NGC 6715 (M54), $\sim$ 0.20 for G1 (Mayall II),  $\sim$ 0.17 for 037 (B327) and $\sim$ 0.20 for B514.

\section{CONCLUSIONS}

Recent discoveries have shown that there is some overlap in
parameter space between the regions occupied by globular clusters and dwarf spheroidal galaxies. It is shown that cluster flattening may be used as an additional parameter to help to distinguish between these two classes of objects. There are, of course, many cases [such as the Fornax dwarf and its globular clusters] where it is quite obvious which object is the dwarf spheroidal and which ones are its companion globular clusters. However, there are other cases where this distinction is not quite so obvious \citep{lee07}. For example, Boo I and Boo II appear to have similar distances and are located at a projected separation of only 1.8 kpc \citet{wal07}. Are these two objects companion galaxies, or is Boo II (M$_{v}$ = -3.1, R$_{h}$ =72 pc) a globular cluster companion to Boo I (M$_{v}$ = -5.7, R$_{h}$ = 227 pc)? Inspection of the outer isophotes of Boo II published by \citet{wal07} suggests that 0.1 
$\la$ (a-b)/a $\la$ 0.2, which is consistent with its being either a galaxy or a cluster.

In summary it is concluded that dwarf spheroidal galaxies are, on
average, significantly flatter than globular clusters. In some cases this may help to distinguish these two classes of objects. Globular clusters that are probably the remnant cores of dwarf spheroidal galaxies appear to have present day flattening values that are, on average, intermediate between these two classes of objects.

\section{ACKNOWLEDGEMENTS}

I am indebted to Bob Abraham, Merla Geha, Bill Harris, Mario Mateo and Shane Walsh for helpful exchanges of e-mail and {\bf to a particular helpful referee}. I also thank Brenda Parrish and Jason Shrivell for technical support.

\begin{table}
\caption[]{Flattening values of Galactic globular clusters and of dwarf nearby spheroidal galaxies}
\begin{tabular}[]{lrr}
\hline
 (a-b)/a  &  n(glob) &  n(dSph)\\
\hline

0.00 - 0.04  &  41    &     1\\
0.05 - 0.09  &  26    &     0\\
0.10 - 0.14  &  16    &     1\\
0.15 - 0.19  &  10    &     0\\
0.20 - 0.24  &   5    &     0\\
0.25 - 0.29  &   2    &     2\\
0.30 - 0.34  &   0    &     5\\
0.35 - 0.39  &   0    &     2\\
0.40 - 0.44  &   0    &     0\\
0.45 - 0.49  &   0    &     0\\
0.50 - 0.54  &   0    &     1\\
0.55 - 0.59  &   0    &     1\\
0.60 - 0.64  &   0    &     0\\
0.65 - 0.69  &   0    &     1\\
0.70 - 0.74  &   0    &     0\\

\hline
\end{tabular}
\end{table}

\begin{figure}
\includegraphics[height=10.5cm]{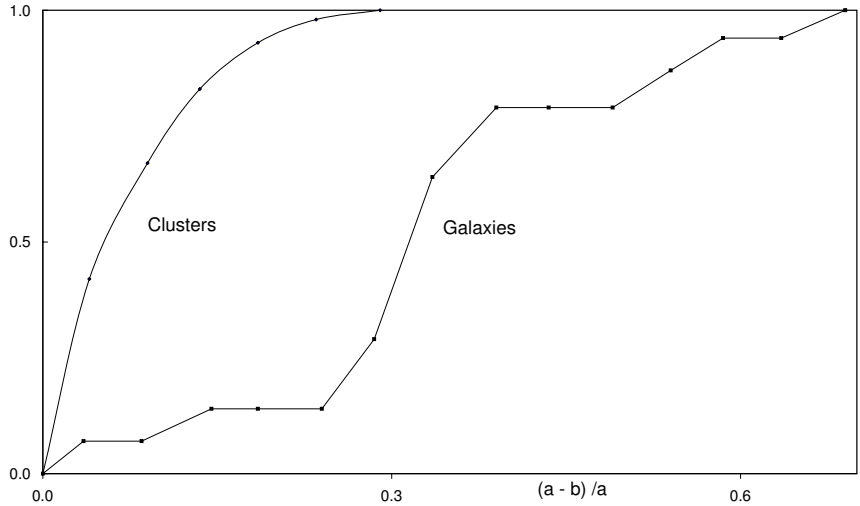}
\caption[]{Normalized cumulative distribution of ellipticity (a-b)/a for Galactic globular clusters compared to that for dwarf spheroidal galaxies within 0.5 Mpc. The figure shows that dwarf spheroidal galaxies are significantly more flattened than are globular clusters.}
\end{figure}

\label{lastpage}
\end{document}